\documentclass{jpp}
\usepackage{graphicx}

\usepackage[utf8]{inputenc}
\usepackage[T1]{fontenc}
\usepackage{amsmath}
\usepackage[colorlinks=true, allcolors=blue]{hyperref}
\usepackage{cleveref}

\newcommand{\Ccool}{C_{\text{cool}}}

\newcommand{\Ccoolbar}{\overline{C}_{\text{cool}}}
\newcommand{\Ccooltilde}{\widetilde{C}_{\text{cool}}}
\newcommand{\tout}{\tau_{\text{outflow}}}
\newcommand{\Bup}{B_{\text{up}}}
\newcommand{\Lsys}{L_{\text{sys}}}
\newcommand{\tcol}{\tau_{\text{collapse}}}
\newcommand{\Bin}{B_\text{in}}
\newcommand{\Bout}{B_\text{out}}
\newcommand{\rhoin}{\rho_\text{in}}
\newcommand{\rhoout}{\rho_\text{out}}
\newcommand{\vin}{v_\text{in}}
\newcommand{\vout}{v_\text{out}}
\newcommand{\nin}{n_\text{in}}
\newcommand{\nout}{n_\text{out}}
\newcommand{\betain}{\beta_{\text{in}}}
\newcommand{\pin}{P_{\text{in}}}

\shorttitle{Current sheet formation under radiative cooling}
\shortauthor{S. Chowdhry, N.F. Loureiro}

\title{Current sheet formation under radiative cooling}

\author{Simran Chowdhry\aff{1}
  \corresp{\email{sc5718@mit.edu}}, Nuno F. Loureiro\aff{1}}

\affiliation{\aff{1}Plasma Science and Fusion Center, Massachusetts Institute of Technology, Cambridge, MA 02139, USA}

\begin{document}

\maketitle

\begin{abstract}
 We present a simple, analytically solvable MHD model of current sheet formation through X-point collapse under optically thin radiative cooling. Our results show that cooling accelerates the collapse of the X-point along the inflows, but strong cooling can arrest or even reverse the current sheet elongation in the outflow direction. Hence, we detail a modification to the radiatively-cooled Sweet-Parker model developed by \citet{uzdensky2011magnetic} to allow for varying current sheet length. The steady-state solution shows that when radiative cooling dominates compressional heating, the current sheet length is shorter than the system size, with an increased reconnection rate compared to the classical Sweet-Parker rate. The model and subsequent results lay out the groundwork for a more complete theoretical understanding of magnetic reconnection in regimes dominated by optically thin radiative cooling.
\end{abstract}

\section{Introduction}
Magnetic reconnection is an ubiquitous process in magnetized plasmas that involves fast, explosive rearrangements of the magnetic field topology, resulting in the conversion of magnetic energy into thermal and kinetic energy \citep{yamada2010magnetic,zweibel2016perspectives,Hesse_Cassak_2020,ji2022magnetic}. 
It drives some of the most energetic events in our solar system such as solar flares \citep{Coppi_Friedland_1971,Priest_1986,Forbes_1991}, coronal mass ejections \citep{Gosling_Birn_Hesse_1995} and geomagnetic storms \citep{Phan_Paschmann_Sonnerup_1996}.
Magnetic reconnection has also been postulated to be responsible for the high-energy radiation observed from many extreme astrophysical environments, such as black hole accretion disks and their coronae \citep{goodman2008reconnection,beloborodov2017radiative,werner2019particle,hakobyan2023radiative}, gamma ray bursts \citep{lyutikov2006electromagnetic,giannios2008prompt,mckinney2012reconnection}, jets from Active Galactic Nuclei \citep{giannios2009fast,petropoulou2023hadronic,nalewajko2014constraining,mehlhaff2020kinetic,sironi2015relativistic}, pulsar wind nebulae \citep{schoeffler2023high,uzdensky2011reconnection_powered,cerutti2012extreme,cerutti2014three,cerutti2017dissipation}, pulsar magnetospheres \citep{lyubarsky2001reconnection,uzdensky2013physical,cerutti2015particle,cerutti2016modelling,philippov2018ab,philippov2019pulsar,hakobyan2019effects,hakobyan2023magnetic} and magnetar magnetospheres \citep{schoeffler2019bright,schoeffler2023high}.

For many of the aforementioned astrophysical environments, the contribution of emitted radiation to the energy partition is significant enough to result in the cooling of the plasma, which in turn is expected to impact reconnection dynamics \citep{oreshina1998slow,somov1976physical,Datta_jpp_2024}. Additionally, remote telescopic observations of radiative emission driven by reconnection are often the only diagnostic probe into the dynamics of these extreme environments \citep{uzdensky2011magneticb}. Thus, a fundamental model of how radiation, particularly radiative cooling, interacts with reconnection processes is an essential part of understanding the microphysics that governs the behavior of these radiation-rich astrophysical plasmas.

The importance of radiatively-cooled reconnection motivated early numerical studies \citep{forbes1991numerical,oreshina1998slow,jaroschek2009radiation}, which show a thinner, denser reconnection layer in the presence of radiative cooling, with slowed-down plasma outflows. More recently, radiative cooling has been included in large-scale PIC simulations of the dynamics of astrophysical environments that are impacted by relativistic reconnection \citep{nalewajko2018kinetic,sironi2020kinetic,sridhar2021comptonization,sridhar2023comptonization}.
Whilst there have been several analytic studies on the interaction of resistive tearing modes and radiation \citep{Steinolfson_1983,Tachi1983,VanHoven1984,Tachi1985}, the first theoretical description of radiatively-cooled reconnection was provided by \cite{uzdensky2011magnetic}, building on previous work by \cite{dorman1995one}. This description extends the Sweet-Parker \citep{parker1957sweet,parker1963solar} model to account for compressibility and predicts the generation of a strongly compressed, colder reconnection layer, which is consistent with results from early simulations \citep{forbes1991numerical,oreshina1998slow,jaroschek2009radiation}. Beyond the Sweet-Parker model, recent simulations \citep{Sen_Keppens_2022,schoeffler2023high,Datta_jpp_2024} and experimental work \citep{Datta_PRL_24} have studied the impact of radiative cooling on reconnection and the plasmoid instability \citep{Loureiro_Uzdensky_2015}, showing the susceptibility of the layer and the formed plasmoids to additional instabilities such as radiative collapse. 

However, this is only a small part of the total picture; the understanding of radiatively-cooled reconnection is in its infancy in comparison to that of reconnection in the classical setting. This paper aims to lay out the groundwork for a more complete theoretical understanding of reconnection in a regime dominated by optically thin radiative cooling by examining the impact of cooling on current sheet formation. This is achieved by revisiting and extending the simple MHD model of current sheet formation through X-point collapse detailed by \citet{Chapman_Kendall_1963} and later \citet{Syrovatskii_1971}, by including a radiative cooling term in the equation of state. The results show that whilst radiative-cooling accelerates the collapse of the X-point along the direction of the inflows, strong cooling can arrest the current sheet elongation in the outflow direction and even result in its reversal and collapse along the outflow direction as well. In context of these results, a steady state solution for radiatively-cooled magnetic reconnection is derived by modifying the model developed by  
\citet{uzdensky2011magnetic} to account for
varying current sheet length. This is achieved by 
enforcing extra condition, requiring the characteristic outflow time to be similar to the cooling time such that a plasma parcel in the layer can be advected out of the layer before undergoing radiatively driven collapse. It is found that for strong radiative cooling that dominates compressional heating, the current sheet length is contracted compared to the system size whilst also yielding an increased reconnection rate as compared to the standard Sweet-Parker model.

This paper is organized as follows; Section \ref{sec:method} details the derivation of the equations dictating radiatively-cooled X-point collapse in the Lagrangian MHD framework, which we adopt for our analytic derivation. Section \ref{sec:results} shows the numerical solutions to the derived equations, analyzes the asymptotic behavior of these solutions and provides a physical interpretation for the same. Section \ref{spmodel} details the calculation used to derive the current sheet length and reconnection rate from our modification of the  radiatively-cooled Sweet-Parker model. Section \ref{discussion} discusses the significance of the results.

\section{Method}\label{sec:method}
We aim to develop a simple model to understand current sheet formation in an ideal MHD plasma that is radiatively cooled by an optically thin cooling mechanism parameterised as $\dot{Q}_{\text{rad}} \sim \rho^aT^b$ \citep{Rybicki_Lightman_2007}, where $\dot{Q}_\text{rad}$ is the volumetric power loss due to radiative cooling, $\rho$ is the plasma mass density and $T$ is the plasma temperature. The plasma is modeled to be compressible, whilst neglecting heat conduction and viscous heating. 
The equations governing this are:
\begin{equation}
    \frac{\partial \rho}{\partial t}+\boldsymbol{\nabla} \cdot(\rho \boldsymbol{v})=0,
    \label{eq:continuity}
\end{equation}
\begin{equation}
\rho\left(\frac{\partial \boldsymbol{v}}{\partial t}+\boldsymbol{v} \cdot \boldsymbol{\nabla} \boldsymbol{v}\right)=-\boldsymbol{\nabla} P+\frac{\boldsymbol{j} \times \boldsymbol{B}}{c}, \label{eq:momentum}
\end{equation}
\begin{equation}
   \frac{\partial \boldsymbol{B}}{\partial t}=\boldsymbol{\nabla} \times(\boldsymbol{v} \times \boldsymbol{B}), 
\label{eq:induction}
\end{equation}
\begin{equation}
  \frac{\partial P}{\partial t}=-(\gamma P \boldsymbol{\nabla}\cdot\boldsymbol{v}+\boldsymbol{v}\cdot\boldsymbol{\nabla}P)-(\gamma-1)C_{\text{cool}}\rho^aT^b ,  
  \label{eq:mhd_eos}
\end{equation}
where the current density $\boldsymbol{j}$ is defined using Amp\'ere's law $\boldsymbol{j}=c(\boldsymbol{\nabla}\times\boldsymbol{B})/4\pi$.
Here, the values of $a$, $b$ and the cooling constant $C_{\text{cool}}$ are specific to the cooling mechanism considered. In the case of thermal bremsstrahlung radiation, $a=2$, $b=1/2$, yielding $C_{\text{cool}}\rho^2T^{1/2}\approx 8.3\times10^{26}\rho^2T^{1/2}[\text{erg s}^{-1}\text{cm}^{-3}]$. \footnote{This value of $C_{\text{cool}}$ can be derived from the exact analytic expression for the bolometric emissivity of thermal bremsstrahlung radiation, $\varepsilon^{f f}=1.4 \times 10^{-27} T^{\frac{1}{2}} Z^2 n_e n_i \bar{g}_B$, where $n_i$ and $n_i$ are the ion and electron number densities, and $\bar{g}_B \approx 1$ is the velocity-averaged Gaunt factor \citep{Rybicki_Lightman_2007}.}

\subsection{Lagrangian MHD} 
\label{sub:lmhd}
Following the formalism laid out by \cite{newcomb62}, as detailed in \cite{Schekochihin_2025}, Lagrangian MHD reformulates the MHD equations into the frame of reference of the fluid parcel. This is particularly useful as it results in a single equation of motion relying solely on the Eulerian fields prescribed at $t=0$.

To derive this Lagrangian equation of motion, we start by defining the relevant frames of reference; $\boldsymbol{r}_0$ is defined as the Lagrangian coordinate with a corresponding Eulerian coordinate $\boldsymbol{r}=\boldsymbol{r}_0+\boldsymbol{\epsilon}(t,\boldsymbol{r}_0,)$. Using this definition along with the known expression for convective derivatives 
$\partial/\partial t_L=d/dt\equiv\partial/\partial t+\boldsymbol{v}\cdot\boldsymbol{\nabla}$, where the subscript `$L$' denotes quantities in the Lagrangian frame, one can then evaluate the velocities,
\begin{equation}
\partial_t \boldsymbol{r}\left(t, \boldsymbol{r}_0\right)=\partial_t \boldsymbol{\epsilon}\left(t, \boldsymbol{r}_0\right) \rightarrow \boldsymbol{v}(t, \boldsymbol{r})=\boldsymbol{v}_L\left(t, \boldsymbol{r}_0\right),
\end{equation}
 and spatial derivatives, 
\begin{equation}
\boldsymbol{\nabla}_0=\left(\boldsymbol{\nabla}_0 \boldsymbol{r}\right) \cdot \nabla=\left(\mathbb{I}+\boldsymbol{\nabla}_0 \boldsymbol{\epsilon}\right)\cdot \boldsymbol{\nabla} . 
\end{equation}

Next, we define the Jacobian relating the Eulerian and Lagrangian frames:
\begin{equation}
    J\left(t,\boldsymbol{r}_0\right)=\left|\operatorname{det} \nabla_0 \boldsymbol{r}\right|=\frac{1}{6} \varepsilon^{i j k} \varepsilon^{l m n} \frac{\partial r_i}{\partial r_{0 l}} \frac{\partial r_j}{\partial r_{0 m}} \frac{\partial r_k}{\partial r_{0 n}} \rightarrow d \boldsymbol{r}=J\left(t,\boldsymbol{r}_0\right) d \boldsymbol{r}_0.
\end{equation}

This expression $J(t,\boldsymbol{r}_0)$, along with the prescribed definitions for velocities and time and spatial derivatives, is everything required to convert all the relevant MHD fields to their Lagrangian counterparts, which will now be denoted using the subscript `$L$'.

As an example, an expression for the Lagrangian mass density $\rho_L(t,\boldsymbol{r}_0)$ can be derived in terms of $\rho_0(\boldsymbol{r}_0)=\rho(t=0,\boldsymbol{r}_0)$ by substituting $J(t,\boldsymbol{r}_0)$ in the continuity equation:
\begin{equation}
\begin{aligned}
        &\rho_0\left(\boldsymbol{r}_0\right) d \boldsymbol{r}_0=\rho(t,\boldsymbol{r}) d \boldsymbol{r}=\rho_L\left(t,\boldsymbol{r}_0\right) d \boldsymbol{r}= \rho_L\left(t,\boldsymbol{r}_0\right) J(t,\boldsymbol{r}_0)d \boldsymbol{r}_0, \\
       & \Rightarrow \rho_L\left(t, \boldsymbol{r}_0\right)=\frac{\rho_0\left(\boldsymbol{r}_0\right)}{J\left(t,\boldsymbol{r}_0\right)}.
\end{aligned}
\end{equation}

Similarly, we can use the induction equation and adiabatic equation of state to obtain corresponding Lagrangian expressions for the magnetic field, $\boldsymbol{B}_L (t,\boldsymbol{r}_0)$,
\begin{equation}
   \boldsymbol{B}_{L}(t,\boldsymbol{r}_0)=\frac{\boldsymbol{B}_0\left(\boldsymbol{r}_0\right) \cdot \nabla_0 \boldsymbol{r}\left(t, \boldsymbol{r}_0\right)}{J\left(t, \boldsymbol{r}_0\right)}, 
\end{equation}

and pressure, $P_L\left(t, \boldsymbol{r}_0\right)$:
\begin{equation}
P_L\left(t, \boldsymbol{r}_0\right)=\frac{P_0}{J^\gamma\left(t, \boldsymbol{r}_0\right)}.
\end{equation}

Plugging all these quantities into the momentum equation yields the following equation of motion, entirely expressed in terms of the Jacobian $J\left(t, \boldsymbol{r}_0\right)$ and initial field values:
\begin{equation}
\label{l_eqm}
\frac{\rho_0}{J} \frac{\partial^2 \boldsymbol{r}}{\partial t^2}=-\left(\boldsymbol{\nabla}_0 \boldsymbol{r}\right)^{-1} \cdot \boldsymbol{\nabla}_0\left(\frac{P_0}{J^\gamma}+\frac{\left|\boldsymbol{B}_0 \cdot \boldsymbol{\nabla}_0 \boldsymbol{r}\right|^2}{8 \pi J^2}\right)+\frac{1}{4 \pi} \frac{\boldsymbol{B}_0}{J} \cdot \boldsymbol{\nabla}_0 \frac{\boldsymbol{B}_0}{J} \cdot \boldsymbol{\nabla}_0 \boldsymbol{r}.
\end{equation}
\subsection{Background: X-Point Collapse}
\label{sec:xpoint_collapse}
Current sheet formation through X-point collapse in an ideal MHD plasma has been extensively studied in the past, with the most notable solutions for an incompressible plasma, as derived by \cite{Chapman_Kendall_1963}, describing the exponential growth of current density $j$. However, to include the impact of radiative cooling, the plasma considered is required to be compressible. This can be understood by considering the instantaneous pressure imbalance introduced due to radiative cooling; cooling results in a drop in thermal pressure, resulting in the compression of the cooled plasma by the inward magnetic pressure as a response. The calculation that includes plasma compressibility and spatially uniform pressure with the generalized equation of state $P=P(\rho)$ was performed by \cite{Syrovatskii_1971}, and will be referred to as S71 for the rest of this paper.

The calculation in S71 is approached in the Eulerian frame of reference with the following initial conditions describing a standard X-point configuration with flows in the $x-y$ plane:
\begin{equation}
\label{xpoint_config}
\begin{gathered}\psi(x, y)=\frac{1}{2}\left(x^2-y^2\right), \\ v_x(x, y)=\Gamma_{x,0} x, \quad v_y(x, y)=\Gamma_{y,0} y, \\ \rho(x, y)=1.\end{gathered}
\end{equation}
Here, the X-point is expressed in terms of the magnetic flux function $\psi$, which has the standard definition $\boldsymbol{B}_0=\boldsymbol{\hat{z}}\times \boldsymbol{\nabla}_{\perp}\psi$ for a background field $\boldsymbol{B}_0$. Dimensionally, $\boldsymbol{B}_0$ has been normalized with respect to the upstream magnetic field $\Bup$ and $x$ and $y$ have been normalized with respect to a characteristic lengthscale `$\Lsys$'. Velocities are normalized with respect to the Alfv\'en velocity $V_A\equiv \Bup/(4\pi\rho_0)^{1/2}$ and times are normalized with respect to the Alfv\'en time $\tau_A\equiv \Lsys/V_A$. Additionally, it is noted that the spatial uniformity of pressure is implied by the condition $\rho(x,y)=1$, which is expressed in normalized units.

For these initial conditions, the MHD equations allow exact similarity solutions, expressed as follows:
\begin{equation}
    \begin{gathered}
        \psi(x, y, t)=\frac{1}{2}\left(\frac{x^2}{\xi^2(t)}-\frac{y^2}{\eta^2(t)}\right), \\ 
        v_x(x, y, t)=\Gamma_x(t) x, \quad v_y(x, y, t)=\Gamma_y(t) y, \\ 
        \rho(x, y, t)=\rho(t).
    \end{gathered}
\end{equation}

Substituting these expressions into the MHD equations above (Equations \ref{eq:continuity}-\ref{eq:mhd_eos} with $\Ccool=0$) yields the following equations relating $\xi$, $\eta$, $\Gamma_x$, $\Gamma_y$ and $\rho$:
\begin{equation}
    \dot{\rho}+\rho (\Gamma_x+\Gamma_y),
    \label{eq:syro_rho}
\end{equation}
\begin{equation}
    \begin{aligned}
        & \rho(\dot{\Gamma}_x+\Gamma_x^2)=-\frac{1}{\xi^2}\left(\frac{1}{\xi^2}-\frac{1}{\eta^2}\right), \\
        & \rho(\dot{\Gamma}_y+\Gamma_y^2)=\frac{1}{\eta^2}\left(\frac{1}{\xi^2}-\frac{1}{\eta^2}\right),     
    \end{aligned}
    \label{eq:syro_gammas}
\end{equation}
\begin{equation}
    \dot{\xi}=\Gamma_x\xi \quad \dot{\eta}=\Gamma_y \eta.
    \label{eq:syro_velocities}
\end{equation}
Thus, combining Equations \ref{eq:syro_rho}, \ref{eq:syro_gammas} and \ref{eq:syro_velocities} yields the following set of equations for $\xi$ and $\eta$: 
\begin{equation}
\label{syro_xi}
    \ddot{\xi}=-\eta\left(\frac{1}{\xi^2}-\frac{1}{\eta^2}\right),
\end{equation}
\begin{equation}
\label{syro_eta}
    \ddot{\eta}=\xi\left(\frac{1}{\xi^2}-\frac{1}{\eta^2}\right),
\end{equation}
with initial conditions $\xi_0=\eta_0=1, \dot{\xi}_0=\Gamma_{x,0}, \dot{\eta}_0=\Gamma_{y,0}$.
The solutions to this set of coupled equations completely prescribe the dynamics of the system.
The numerical solutions (with $\Gamma_x<0$ and $\Gamma_y>0$), as seen in Figure \ref{fig:eta_xi_sol}, show that $\eta$ grows over time, while $\xi$ collapses and exhibits a finite-time singularity, representing a forming current sheet with inflows and outflows along the $x$- and $y$-axes, respectively, and a diverging current density $j=(1/\xi^2-1/\eta^2)\to \infty$ (obviously, in a realistic system resistivity would eventually become important and curb the growth of the current).

When performing this calculation in Lagrangian MHD, it can be seen that the equations for $\xi$ and $\eta$ can easily be recovered by considering the self-similar solution as a coordinate conversion $x=\xi(t) x_0, \quad y=\eta(t) y_0, \quad z=z_0$ that is substituted into Equation \ref{l_eqm}. Using this coordinate conversion, the following expressions can be obtained for $J$, 
\begin{equation}
    J\left(t,\boldsymbol{r}_0\right)=\left|\operatorname{det} \nabla_0 \boldsymbol{r}\right|=\frac{1}{6} \varepsilon^{i j k} \varepsilon^{l m n} \frac{\partial r_i}{\partial r_{0 l}} \frac{\partial r_j}{\partial r_{0 m}} \frac{\partial r_k}{\partial r_{0 n}}=\xi(t)\eta(t),
    \label{eq:J}
\end{equation}
and for $\boldsymbol{\nabla}_0 \boldsymbol{r}$ and $\left(\boldsymbol{\nabla}_0 \boldsymbol{r}\right)^{-1}$:
\begin{equation}
    \boldsymbol{\nabla}_0 \boldsymbol{r}=\begin{bmatrix}
\xi &0\\ 0 & \eta
\end{bmatrix}, \left(\boldsymbol{\nabla}_0 \boldsymbol{r}\right)^{-1}=\begin{bmatrix}
1/\xi &0\\ 0 & 1/\eta
\end{bmatrix}.
\label{eq:lagrange_grads}
\end{equation}
Substituting these expressions into Equation \ref{l_eqm}  exactly yields Equations \ref{syro_xi} and \ref{syro_eta}.
\subsection{Lagrangian Equation of Motion with Radiative Cooling}
\label{sub:leom_rc}

Sticking to the Lagrangian formalism, we aim to find modified equations for $\xi$ and $\eta$ which include the effect of optically thin radiative cooling, by keeping the entire right-hand side of the equation of state (Equation \ref{eq:mhd_eos})
This implies that whilst the coordinate conversion of density and magnetic field remain the same, the conversion of pressure changes such that it takes a form that satisfies the modified equation of state. To find this, we first recast the equation of state to isolate the cooling term and express it entirely in terms of pressure $P$ and 
 mass density $\rho$: 
\begin{equation}
\frac{d}{d t} \ln \left[\frac{P}{\rho^\gamma}\right]=-(\gamma-1) C_{\text{cool}}\left[\frac{(Z+1)}{m_i+Zm_e}\right]^{-b} \rho^{a-b} P^{b-1}.
\end{equation}
This additional factor of $[(Z+1)/(m_i+Zm_e)]^{-b}$ comes from converting number density $n$ to mass density $\rho$.

Defining
\begin{equation}
\Ccoolbar \equiv(\gamma-1) C_{\text{cool}}\left[\frac{(Z+1)}{m_i+Zm_e}\right]^{-b},
\end{equation}
the corresponding Lagrangian equation of state is:
\begin{equation}
\frac{\partial}{\partial t} \ln \left[\frac{P_L}{\rho_L^\gamma}\right]=-\Ccoolbar \rho_L^{a-b} P_L^{b-1}.
\label{eq:lagrangian_eos}
\end{equation}
Given that $P_L\rho_L^{-\gamma}=P_0\rho_0^{-\gamma}$ at $t=0$, we can use the ansatz $P_L\rho_L^{-\gamma}=F(t,x,y)P_0\rho_0^{-\gamma}$, with $F(t=0,x,y)=1$, in the equation of state to obtain the following general solution for $P_L$ (solved in Appendix \ref{appA}):
\begin{equation}
P_L=\frac{P_0}{J^\gamma} \left(C_1-\frac{C_2}{P_0}\left[(\gamma-1)(1-b) \Ccoolbar \int_0^t J^{b(1-\gamma)+(\gamma-a)}\left(t',\boldsymbol{r}_0\right) d t^{\prime}\right]^{1 /(1-b)}\right).
\label{eq:lagrangian_pressure}
\end{equation}
Given that the initial conditions can entirely be satisfied by setting $C_1=1$, the choice of $C_2(x_0,y_0)$ is unconstrained. It can be seen that $C_2(x_0,y_0)$ is dimensionally a pressure term, thus, a sensible choice would be to set it to $P_0(x_0,y_0)$. To fully specify this solution, a pressure profile needs to be selected; for this, we choose to set a parabolic pressure profile
\begin{equation}
    P_0\left(x_0, y_0\right)=P_{0, X}\left(1-\frac{1}{2}\left(x_0^2+y_0^2\right)\right),
    \label{eq:para_profile}
\end{equation}
where $P_{0,X}$ is a constant signifying the maximum magnitude of the pressure. This is physically similar to what would be expected in a reconnection layer, with the pressure and heating concentrated near the current sheet.
With these choices, Equation \ref{eq:lagrangian_pressure} yields the following final form for $P_L$:
\begin{equation}
    P_L=\frac{P_0\left(x_0, y_0\right)}{J^\gamma}\left(1-\Ccooltilde\left[ \int_0^t J^{b(1-\gamma)+(\gamma-a)}\left(t',\boldsymbol{r}_0\right) d t^{\prime}\right]^{1 /(1-b)}\right),
\end{equation}
with all the relevant cooling constants absorbed into $\Ccooltilde$:
\begin{equation}
  \Ccooltilde=[(\gamma-1)(1-b)\Ccoolbar]^{1/(1-b)}.
  \label{eq:ctilecool}
\end{equation}

Plugging this into the momentum equation along with the standard definitions for $B_L$ and $\rho_L$ yields the following modified Lagrangian equation of motion:
\begin{equation}
\label{l_rcool_eqm}
\frac{\rho_0}{J} \frac{\partial^2 \boldsymbol{r}}{\partial t^2}=-\left(\boldsymbol{\nabla}_0 \boldsymbol{r}\right)^{-1} \cdot \boldsymbol{\nabla}_0\left(\frac{P_0}{J^\gamma}\left[\Ccooltilde I[J]^{1/(1-b)}-1\right]+\frac{\left|\boldsymbol{B}_0 \cdot \boldsymbol{\nabla}_0 \boldsymbol{r}\right|^2}{8 \pi J^2}\right)+\frac{1}{4 \pi} \frac{\boldsymbol{B}_0}{J} \cdot \boldsymbol{\nabla}_0 \frac{\boldsymbol{B}_0}{J} \cdot \boldsymbol{\nabla}_0 \boldsymbol{r},
\end{equation}
where:
\begin{equation}
  I[J]=\int_0^t J^{b(1-\gamma)+(\gamma-a)}\left(t',\boldsymbol{r}_0\right) d t^{\prime}.  
\end{equation}

This expression notably differs from that in Equation \ref{l_eqm} by including an expression for pressure that is modified by radiative cooling and a non-uniform initial pressure by a factor of $\Ccooltilde I[J]^{1/(1-b)}-1 $.

Starting with the same X-point configuration as S71 (Equations \ref{xpoint_config}), a Lagrangian equation of motion that includes radiative cooling can be obtained by plugging in the coordinate conversion $x=\xi(t) x_0, \quad y=\eta(t) y_0, \quad z=z_0$. Similar to Section \ref{sec:xpoint_collapse}, substituting the expression for J (Equation \ref{eq:J}) and and $\boldsymbol{\nabla}_0 \boldsymbol{r}$ and $\left(\boldsymbol{\nabla}_0 \boldsymbol{r}\right)^{-1}$ (Equations \ref{eq:lagrange_grads}) into Equation \ref{l_rcool_eqm} yields the following matrix equation for $\xi$ and $\eta$:

\begin{equation}
\frac{1}{\xi\eta} \begin{bmatrix}
    x_0\ddot{\xi}\\
    y_0\ddot{\eta}
\end{bmatrix}
=-\frac{(\Ccooltilde I[\xi\eta]^{1/(1-b)}-1)}{(\xi\eta)^\gamma} \begin{bmatrix}
    \partial_{x_0}P_{0}/\xi \\
    \partial_{y_0}P_{0}/\eta
\end{bmatrix}
+ \begin{bmatrix}
    x_0/\xi^3 \\
    y_0/\eta^3
\end{bmatrix}
+\frac{1}{\xi\eta} \begin{bmatrix}
    x_0/\eta \\
    y_0/\xi
\end{bmatrix}.
\end{equation}
 Finally, a set of spatially independent coupled equations for $\xi$ and $\eta$ can be found using the selected form of $P_0\left(x_0, y_0\right)$ in Equation \ref{eq:para_profile}:
\begin{equation}
\label{final_rcool_eq}
\begin{aligned}
& \ddot{\xi}=-\eta\left[\frac{1}{\xi^2}-\frac{1}{\eta^2}+\frac{P_{0,X}(\Ccooltilde\left(I[\xi \eta]\right)^{\frac{1}{1-b}}-1)}{(\xi \eta)^\gamma}\right], \\
& \ddot{\eta}=\xi\left[\frac{1}{\xi^2}-\frac{1}{\eta^2}-\frac{P_{0,X}(\Ccooltilde\left(I[\xi \eta]\right)^{\frac{1}{1-b}}-1)}{(\xi \eta)^\gamma}\right].
\end{aligned}
\end{equation}
The equations derived in S71 (Equations \ref{syro_xi} and \ref{syro_eta}) can be recovered by setting $\Ccooltilde=0$ (no radiative cooling) and setting $P_0$ to be uniform. Looking at the form of these equations, we can qualitatively anticipate what the impact of radiation would be. First considering the equation for $\ddot{\xi}$; given that the $1/\xi^2$ terms and cooling terms have the same sign, radiative cooling contributes to $\ddot{\xi}$ becoming more negative, implying that radiation accelerates the collapse of $\xi$. In the equation for $\ddot{\eta}$, the $1/\xi^2$ term and the cooling term have opposite signs; $\ddot{\eta}$ can become negative if the cooling term dominates over $1/{\xi^2}$, resulting in the deceleration of the outflows.

To find a reasonable physical interpretation for $\Ccooltilde$, we re-express this term as a function of the cooling parameter:
\begin{equation}
R\equiv\frac{\tout}{\tau_{\text{cool}}}.
\end{equation}
Here, $\tau_{\text{cool}}$ is the effective cooling time:
\begin{equation}
    \tau_{\text {cool }} \equiv \frac{P}{\dot{Q}_{\text {rad }}} \approx \frac{P^{1-b}\rho^{b-a}}{\Ccoolbar},
    \label{eq:tau_cool}
\end{equation}
and $\tout$ is the timescale corresponding to the advection of the outflows:
\begin{equation}
    \tout =\frac{\eta}{v_y(y=\eta)}=\frac{1}{\Gamma_y}
    \label{eq:tau_alfven},
\end{equation}
where $\eta$ is the characteristic length of the current sheet at a given time with a corresponding outflow velocity $\Gamma_y\eta$.
$R$ is directly related to the strength of radiative cooling as it is a measure of how fast a fluid parcel can be radiatively cooled before it is advected away. 
For the configuration in this paper, the general definitions of $\tau_{\text{cool}}$ and $\tout$ are expected to be non-uniform and time dependent, thus $\tau_{\text{cool}}$ and $\tout$ in the expression for $R$ need to be defined at a specific time and position. We choose to define $R$ at $t=0$ at the X-point, yielding:
\begin{equation}
R=\Ccoolbar \Gamma_{y,0}P_{0,X}^{b-1}\rho_0^{a-b}.
\end{equation}
By rearranging this expression to isolate $\Ccoolbar$ and substituting it in Equation \ref{eq:ctilecool}, the following relationship between $\Ccooltilde$ and $R$ can be obtained:
\begin{equation}
 \Ccooltilde=\left[\frac{(\gamma-1)(b-1)}{\rho_0^{a-b}P_{0,X}^{b-1}\Gamma_{y,0}}R\right]^{\frac{1}{1-b}}.
    \label{eq:ct_in_rc}
\end{equation}

More generally, $R$ and $\Ccooltilde$ can be physically understood by considering the relative strength of radiative cooling compared to compressional heating in Equation \ref{eq:mhd_eos}. 
First, we define the ratio between the characteristic magnitudes of compressional heating and radiative cooling using a dimensionless variable $\mathcal{R}$:
\begin{equation}
    \mathcal{R}\equiv \frac{|(\gamma-1)\Ccool\rho^aT^b|}{|\gamma P \boldsymbol{\nabla}\cdot\boldsymbol{v}|} = \frac{\Ccoolbar\rho^{a-b}P^b}{\gamma P c_s/\Lsys},
    \label{eq:fancyR}
\end{equation}
where $c_s$ is the sound speed of the ambient plasma $u=c_s=(\gamma P_0/\rho_0)^{1/2}$. 
Substituting the expression for $\mathcal{R}$ into Equation \ref{eq:tau_cool}, and estimating the outflows to be Alfv\'enic yields the following simplified expression for $R$:
\begin{equation}
    R=\frac{\gamma c_s}{V_A}\mathcal{R}=\gamma^{1/2}\beta^{1/2}\mathcal{R},
\end{equation}
where $\beta$ is the ratio between the plasma pressure and the magnetic pressure.
Thus, $R$ and by extension $\Ccooltilde$ is measure of the impact of radiative cooling over counteracting plasma compression effects (heating through $\mathcal{R}$ and the pressure imbalance through $\beta$).

\section{Results and Analysis} 
\label{sec:results}
In this section, we show the numerical solutions to the equations for $\ddot{\xi}$ and $\ddot{\eta}$ in Equation \ref{final_rcool_eq} and analyze their asymptotic behavior for cases with both weak and strong radiative cooling.
\subsection{Numerical Solutions}
\label{sub:numerical}
For selected values of $\Ccooltilde$ (Equation \ref{eq:ct_in_rc}), Equations \ref{final_rcool_eq} can be solved numerically to obtain solutions for the location of a plasma parcel initially at $(x_0,y_0)$ in terms of $\xi$ and $\eta$, and by extension, velocity in terms of $\Gamma_x=\dot{\xi}/\xi$ and $\Gamma_y=\dot{\eta}/\eta$.
We choose to set the values of $\xi$ and $\eta$ at $t=0$ to be $\xi_0=1$ and $\eta_0=1$ to match the configuration used in S71. Additionally, the chosen geometry of the current sheet is such that the outflows are along the $y$-axis and the inflows along the $x$-axis; this requires $\Gamma_{y,0}>0$ and $\Gamma_{x,0}<0$.

\begin{figure}
    \centering
    \includegraphics[page=1,width=1.0\textwidth]{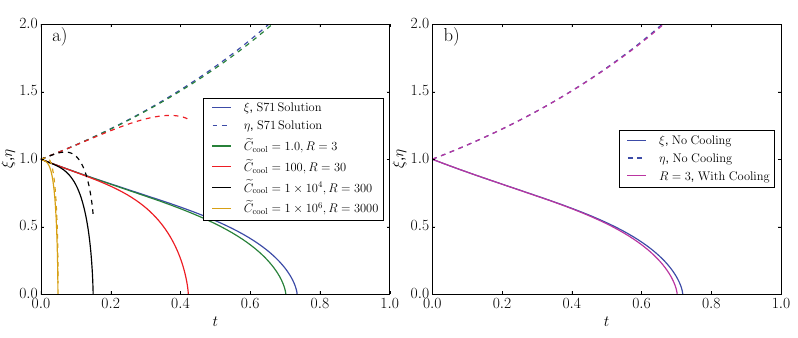}
    \caption{a) Plot of $\xi$ (solid) and $\eta$ (dashed) vs time `t' for Syrovatskii's solution ($R=0$, blue), $R=3$ (weak cooling, green), $R=30$ (red), $R=300$ (black) and $R=3000$ (strong cooling, yellow). It can be seen that stronger radiative cooling causes the X-point to collapse more rapidly in the inflow direction, and also potentially results in a reversal of the direction of the outflows.  b) Plot comparing the impact of non-uniform pressure without radiative cooling (blue) and with weak radiative cooling ($R=3$, pink) on $\xi$ (solid) and $\eta$ (dashed). For the same initial magnitude, the pressure modified by radiative cooling has a smaller $t_c$; thus the effects observed in a) can be attributed to radiative cooling, rather than just the non-uniform pressure.}
    \label{fig:eta_xi_sol}
\end{figure}

Next, realistic values of $\Ccooltilde$ that are astrophysically relevant need to be selected. To achieve this, we choose four values of $R$; 3 (weak cooling), 30, 300 and 3000 (strong cooling), which correspond to $\Ccooltilde$ values of $1$, $100$, $1\times10^4$ and $1\times 10^6$ respectively. These values represent astrophysical environments ranging from the solar corona, which is dominated by line radiation with $R\approx 3$ \citep{Ji_Daughton_2011,Sen_Keppens_2022} to the magnetosphere of the Crab Nebula, which is strongly cooled by synchrotron radiation with $R\approx 3000$ \citep{Uzdensky_Spitkovsky_2013},

In Figure \ref{fig:eta_xi_sol}(a), 
it can be seen that initially, close to $t=0$, $\xi$ and $\eta$ evolve similarly for all values $R$, with $\xi$ displaying linear collapse and $\eta$ displaying linear growth. Additionally, it can be seen that irrespective of the strength of radiative cooling, there exists a finite-time singularity in the solution for $\xi$. However, in the cases with stronger radiative cooling, it can be seen that the time at which $\xi$ becomes singular, which we define as the critical time $t_c$, is reached earlier, implying that stronger cooling results in faster collapse of the X-point.
Another particularly interesting feature that can be seen in the cases with stronger cooling, particularly $R=300$ (black) and $R=3000$ (yellow) is the presence of a turning point in the solution for $\eta$, implying that for strong enough cooling, the outflows reverse before the X-point collapses entirely.

\begin{figure}
    \centering
    \includegraphics[page=2,width=1.0\textwidth]{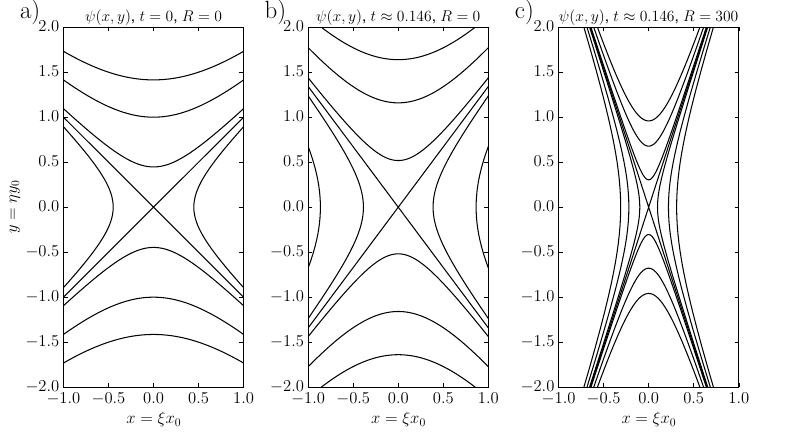}
    \caption{a) Plot of the initial X-point configuration in terms of $\psi$, where $\xi=\xi_0=1$ and $\eta=\eta_0=1$ b) and c) Plot of the same set of contours of $\psi$ as part (a) with no radiative cooling (b) for $t=0.146s$, which is near the critical time $t_c=0.15s$ for the radiatively-cooled case with $R=300$ (c). It can be seen that whilst both cases have a narrower X-point in the inflow direction, the radiatively-cooled X-point is significantly narrower in the inflow direction and also appears to contract in the outflow direction, which is indicative of the flow reversal observed in Figure \ref{fig:eta_xi_sol}.}
    \label{fig:xpointevolution}
\end{figure}

We can compare how the structure of the X-point is impacted by radiative cooling using 2-D contour plots of the magnetic flux function $\psi$. In Figure \ref{fig:xpointevolution}, we plot $\psi(x,y)$ at $t=0$ (a), and $t=0.146$ for $R=0$ (b) and $R=300$ (c). The time $t=0.146$ has been selected as it is close to $t_c$ for $R=300$, showing a clear contrast between the radiatively-cooled and non-radiative case. 
As expected, both the non-radiative and radiatively-cooled cases have a narrower X-point along the inflow direction in comparison to the configuration (Figure \ref{fig:xpointevolution}(a). However,  the radiatively-cooled X-point is significantly narrower that the non-radiative X-point at the same point in time; this is consistent with the results shown in Figure \ref{fig:eta_xi_sol}(a), implying that radiative cooling results in faster collapse in the inflow direction. 
Additionally, it can also be seen that the radiatively-cooled X-point is also collapsing along the outflow direction, unlike the non-radiative X-point, which continues to elongate; this is consistent with the the flow reversal implied by the eventual decrease in $\eta(t)$ observed in Figure \ref{fig:eta_xi_sol}(a).

Lastly, in Figure \ref{fig:eta_xi_sol}(b), we compare the impacts of the simple non-uniform pressure term (blue) and the pressure term modified by radiative cooling (pink) on the evolution of $\xi$ and $\eta$. A crucial difference in the physics in this paper compared to the derivation in S71 is the inclusion of non-uniform pressure; the derivation in S71 assumes the pressure to be uniform and thus not included in dynamics of $\xi$ and $\eta$. It can be seen as the solution with radiative cooling included approaches the finite-time singularity earlier, the significant impact on the solution of $\eta$ as well as $t_c$ can primarily be attributed to radiative cooling.

\subsection{Asymptotic Behavior near Finite-Time Singularity}
\label{subsec:asymptotics}

\begin{figure}
    \centering
    \includegraphics[page=3,width=1.0\textwidth]{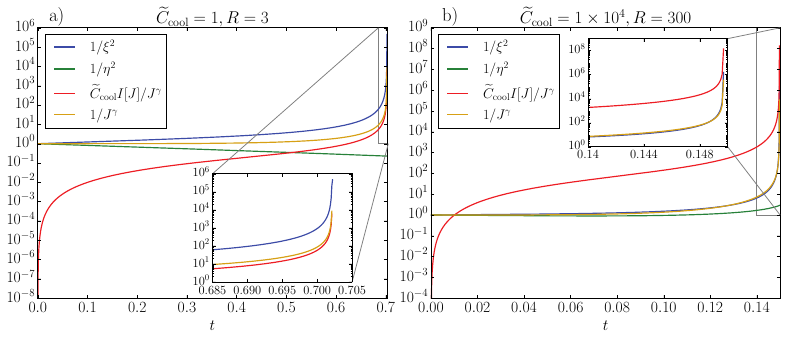}
    \caption{a) Plot of each term in the equation for $\ddot{\xi}$ (Equation \ref{l_rcool_eqm}) for $R=3$ (weak cooling) up to $t=t_c$. It can be seen that whilst $1/\xi^2$ (blue) is the dominant term, with a small contribution from radiative cooling (red). b) Plot for $R=300$ (strong cooling) up to $t=t_c$, clearly showing radiative cooling (red) to be the dominant term. In both plots, it can be seen that the contribution of $1/\eta^2$ (green) is negligible, whilst the albeit more significant contribution of the $1/J^{\gamma}$ term (non-uniform pressure, yellow) can be neglected near $t=t_c$ }
    \label{fig:individual_terms}
\end{figure}

The numerical solutions for both the case in S71 and the radiatively-cooled case show the presence of a finite-time singularity at $t=t_c$ where $\xi=0$. The behaviour $\xi$ for $t\approx t_c$ can be analysed analytically as in this region, $1/\xi^2$ is large and $\eta \sim \eta (t_c)$.  This is done by Syrovatiskii in S71 (using Equation \ref{syro_xi}) to yield the scalings: \begin{equation}
  \begin{aligned} & \xi \sim\left(t_c-t\right)^{2/3}, \eta \sim \eta\left(t_c\right) \rightarrow \gamma \sim-\left(t_c-t\right)^{-1}, \delta=\text { constant } \\ & \Rightarrow \rho \sim\left(t_c-t\right)^{-2/3}, j \sim\left(t_c-t\right)^{-4/3}\end{aligned}  
  \label{eq:syro_scale}
\end{equation}
Applying the same analysis for the radiatively-cooled case is non-trivial as the treatment of the integral containing the radiative cooling term is not obvious. Thus, we consider the following two different limiting cases:
\begin{enumerate}
\item Small $R$, weak radiative cooling: Here, we can consider the radiative cooling term as a small correction to S71. The integral in this term can be evaluated using the non-radiative scaling $\xi \sim (t_c-t)^{2/3}$, allowing us to solve for $\xi$ pertubatively.
\item Large $R$, strong radiative cooling: Here, we substitute the algebraically parameterized ansatz $\sim (t_c-t)^{\alpha}$ into the equation for $\xi$ and assume a dominant balance between $\ddot{\xi}$ and the radiatively-cooled term.
\end{enumerate}
The dominant balances and methodology used for each case are confirmed using the plots in Figure \ref{fig:individual_terms}, which compares the size of each term in the equation for $\ddot{\xi}$ in Equation \ref{final_rcool_eq}.
\subsubsection{Weak Cooling: Small $R$ Corrections}
First considering the case with small $R$; as before, near $t\approx t_c$, $1/\xi^2\gg1/\eta^2$, allowing us to neglect the $1/\eta^2$ term. Additionally, as $\eta\approx \eta(t_c)\gg\xi$, we also consider the $1/(\xi\eta^{\gamma})$ term to be negligible given that $\gamma=5/3<2 \to 1/\xi^2\gg1/(\eta(t_c)\xi)^{5/3}$. Substituting $\xi\sim(t_c-t)^{2/3}$ and $\eta\approx \eta(t_c)$ into $I[\xi\eta]$ thus gives the following expression for $\ddot{\xi}$
\begin{equation}
    \ddot{\xi} \approx-\eta\left(t_c\right)\left[\frac{1}{\xi^2}+C_{\text{int}}(t_c-t)^{2\alpha_p/3+1/(1-b)}\right] \quad \text{and} \quad \alpha_p=\frac{b-a}{(1-b)},
    \label{eq:diffeq_xi_smallrc}
\end{equation}
where:
\begin{equation}
    C_{\text{int}}=\Ccooltilde \left[\frac{9}{2} \eta(t_c)^4\right]^{(b-a)/ 3(1-b)}\left[\frac{3}{2[b(1-\gamma)+(\gamma-a)]+3}\right]^{1/(1-b)}.
\end{equation}

For bremsstrahlung cooling, with $a=2,b=1/2$ and $\gamma=5/3$, this equation becomes:
\begin{equation}
    \ddot{\xi} \approx -\eta(t_c)\left(\frac{1}{\xi^2}+C_{\text{int}}\right).
\end{equation}
Using the following labeling of constants:
\begin{equation}
\begin{aligned}
& c_1=-\eta(t_c), \\
& c_2=-C_{\text{int}} \eta(t_c), \\
& c_3=\frac{c_1^{-1 / 3}/2-c_2c_1^{2 / 3}}{c_1^{1 / 3}}, \\
&\xi(t=t_c)=0,
\end{aligned}
\end{equation}
the general solution to this equation (solved in Appendix \ref{appB}) can be written as:
\begin{equation}
    (t_c-t)=\int^{\xi(t_c)}_{\xi_0} \sqrt{\frac{\xi}{\left(2 c_3 c_1^{2 / 3} \xi+2 c_2 \xi^2-2 c_1\right)}} d \xi.
\end{equation}
For small $\xi$, the integrand is $\sim O(\xi^{1/2})$ and thus the dominant contribution of this integral comes from the region where $t\approx t_c$ to yield:
\begin{equation}
    \left(t_c-t\right)=\frac{2}{3} \sqrt{-\frac{1}{c_1}} \xi^{3 / 2}-\frac{2}{5}\left[c_3 c_1^{2/3}\left(-\frac{1}{c_1}\right)^{3/2}\right]\xi^{5/2}+O(\xi^{7/2}).
\end{equation}
To lowest order, this retrieves the S71 scaling of $\xi \sim (t_c-t)^{2/3}$. When plugging in this scaling with a corrective term to $\sim O(\xi^{5/2})$, we find a correction term which scales as $\sim (t_c-t)^{4/3}$.
This scaling (see Appendix \ref{appB} for the full expression) is a good fit for the data, as shown in Figure \ref{fig:scaling_fit}(a), where the curve representing this scaling (blue) converges to the data (green) as $t$ approaches $t_c$.

\subsubsection{Strong Cooling: Algebraic Parameterization}
Now, the case where $R$ is large can be analyzed by parameterizing $\xi$ near $t_c$ as $\sim (t_c-t)^{\alpha}$ and solving for $\alpha$ in the equation for $\ddot{\xi}$ (Equation \ref{final_rcool_eq}) by considering the appropriate dominant balances:
\begin{equation}
    \ddot{\xi} \approx - \eta\left(t_c\right)\left[\left(t_c-t\right)^{-2 \alpha}+\frac{C_{\text{int}}\left(t_c-t\right)^{\frac{\alpha[b(1-\gamma)+(\gamma-a)]+1}{(1-b)}}}{\eta\left(t_c\right)^\gamma\left(t_c-t\right)^{\alpha\gamma}}\right],
\end{equation}

\begin{equation}
    (t_c-t)^{\alpha-2}\approx \eta\left(t_c\right)\left[\left(t_c-t\right)^{-2 \alpha}+\frac{C_{\text{int}}\left(t_c-t\right)^{\frac{\alpha(b-a)+1}{(1-b)}}}{\eta\left(t_c\right)^\gamma}\right].
\end{equation}
There are two reasonable dominant balances here based on the regime. First, with no radiative cooling, the dominant balance is
\begin{equation}
  (t_c-t)^{\alpha-2}\sim (t_c-t)^{-2\alpha} \to \alpha=\frac{2}{3}, 
\end{equation}
which returns the scaling in S71, as expected. 
The other dominant balance, which is evident for a sufficiently large $R$, as seen in Figure \ref{fig:individual_terms}(b), is:
\begin{equation}\ddot{\xi}\sim\frac{C_{\text{int}}\left(t_c-t\right)^{\frac{\alpha(b-a)+1}{(1-b)}}}{\eta\left(t_c\right)^\gamma} \to (t_c-t)^{\alpha-2}\sim (t_c-t)^{\frac{\alpha(b-a)+1}{(1-b)}} \to \alpha\approx\left[\frac{3-2b}{1-2b+a}\right]. 
\end{equation}
For bremsstrahlung cooling, this gives $\alpha \approx1$, yielding:
\begin{equation}
    \xi \sim (t_c-t), \eta \approx \eta(t_c)
    \label{eq:scale_largerc}.
\end{equation}
Here, we still assume that $\eta \ll \xi$ near $t=t_c$; though technically there is a large enough value of $R$ such that $\xi \sim \eta$. However, we are not aware of any astrophysical environment where such a large value of $R$ would be relevant.

\begin{figure}
    \centering
    \includegraphics[page=4,width=1.0\textwidth]{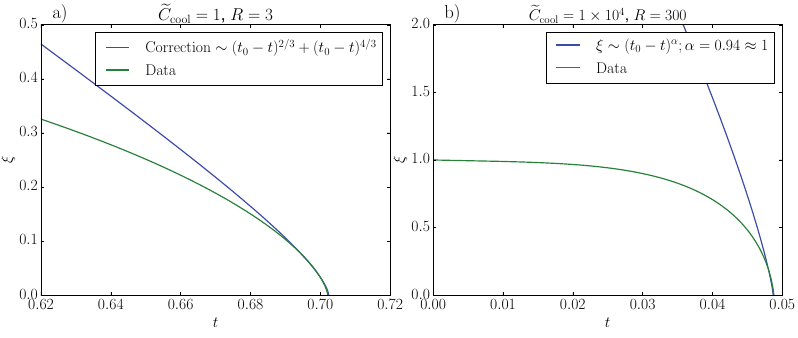}
    \caption{Plots comparing the derived scalings near the finite-time singularity (blue) to the numerical solution for $\xi$ (green) for $R=3$ (a, weak cooling) and $R=300$(b, strong cooling). Here, the range in $t$ has been selected to highlight behaviour near $t\approx t_c$.}
    \label{fig:scaling_fit}
\end{figure}

Similar to the scaling for small $R$, this scaling is shown to be a good fit for the data, in Figure \ref{fig:scaling_fit}(b).

Given the results obtained for weak cooling in Equation \ref{eq:weakcool_scale} and strong cooling in Equation \ref{eq:scale_largerc}, it can be seen that radiative cooling accelerates the collapse of $\xi$ near $t\approx t_c$, with stronger cooling resulting in more rapid collapse.

\subsection{Force Balance and Critical $R$}
\label{sub:rcrit}
A particularly prominent feature in the solutions for $\xi$ and $\eta$ in Figure \ref{fig:eta_xi_sol}(a) is the presence of a turning point in $\eta$, implying the reversal of the outflows in the presence of strong radiative cooling (higher values of $R$). 
Given this result, we would like to understand what is the underlying physical mechanism that dictates such reversal, and what is the critical value of $\Ccooltilde$, and thus $R$, above which this occurs.

The underlying physical mechanism dictating this reversal of flows can be qualitatively understood by separately considering the pressure imbalance due to radiative cooling and the subsequent response along the $x$-axis and the $y$-axis. Starting with the initial configuration of dominant forces due to pressure gradients; along the $y$-axis (outflow direction), the force on a plasma parcel due to both the thermal pressure gradient and the $\mathbf{j}\times \mathbf{B}$ force point in the direction of the outflows, yielding a positive kinetic pressure gradient. This is because the thermal pressure profile is parabolic with its maximum at the X-point, and the outflow velocity is positive ($\Gamma_{y,0}>0$). Along the $x$-axis, the force on the plasma parcel due to the magnetic pressure gradient points in the direction of the inflows and is opposed by the force due to the thermal pressure gradient.  Dynamically, radiative cooling (which is stronger at $y_0=0$ due to the parabolic pressure profile) results in a drop in temperature and thus thermal pressure. This results in an instantaneous increase in pressure imbalance in the $x$-direction, resulting in the compression of the plasma and subsequent increase in density by the magnetic pressure. Whilst both thermal pressure and the radiative cooling power density $\dot{Q} \sim \rho^aT^b$ increase with an increase in density, cooling mechanisms with $a>1$ will result in an additional drop in thermal pressure. Notably, this runaway drop in thermal pressure is expected to be more significant at the X-point as the radiative cooling is stronger at $(x_0, y_0)=(0,0)$ due to the parabolic pressure profile, resulting in stronger compression at the X-point by the magnetic pressure. As a result, the thermal pressure at the X-point will eventually fall below that of the surrounding plasma, reversing the direction of the thermal pressure gradient. The force induced due to this therefore opposes $\mathbf{j}\times \mathbf{B}$ force that drives the kinetic pressure gradient of the outflows. For strong enough cooling (large enough $R$), this inward pointing thermal pressure gradient can become large enough to not only balance but dominate the opposing forces, consequently stalling and reversing the direction of the outflows.

We choose to define the critical $R\equiv R^{\star}$ to be the value of $R$ at which the cooling term is of the same order as $1/\xi^2$ near the finite-time singularity; this implies that the cooling term at the very least balances other terms at its maximum, which it reaches at $t=t_c$. 
With the natural assumption that the observed impact on $\
\eta$ can be attributed to radiative cooling, we would expect radiative cooling mechanisms with values of $R$ distinctly greater than $R^{\star}$ to result in the deceleration and reversal of the outflows. 

From the analysis of the behavior of $\xi$ near the finite-time singularity and the results in Figure \ref{fig:eta_xi_sol}(a), $R^{\star}$ likely inhabits an intermediate cooling regime, for which the behavior of $\xi(t)$ near $t_c$ is unknown. However, it can be assumed that for such a regime,  the following terms in the equation for $\ddot{\xi}$ (Equation \ref{final_rcool_eq}) balance:
\begin{equation}
  \frac{\ddot{\xi}}{\eta(t_c)} \sim \frac{1}{\xi^2}\sim \frac{\Ccooltilde I[\xi\eta(t_c)]^{1/(1-b)}}{\eta(t_c)^{\gamma}\xi^\gamma},
  \label{eq:rstar_balance}
\end{equation}
and thus have the same time dependence near $t_c$, with an integration constant of order unity. It can be noted that as 
there is no expected turning point in $\eta$ for $t\ll t_c$ in this intermediate regime, $\xi\ll\eta$, allowing us to neglect the $1/(\xi\eta)^{\gamma}$ and $1/\eta^2$ terms. Thus, considering the average value of $\eta(t_c)$ at that point (see Figure \ref{fig:eta_xi_sol}) along with the integration constants for bremsstrahlung cooling, $\Ccooltilde^{\star}$ can be estimated from Equation \ref{eq:rstar_balance} to be $\Ccooltilde^{\star} \approx 10$, with a corresponding $R^{\star}\approx 10$.

\begin{figure}
    \centering
    \includegraphics[page=5,width=1.0\textwidth]{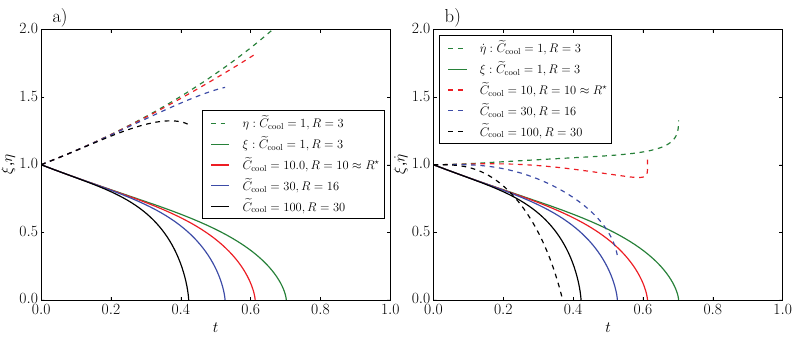}
    \caption{a) Plot of $\xi$ (solid) and $\eta$ (dashed) vs time $t$ for $R=3$ (green), $R=10 \approx R^{\star}$ (blue), $R=16$ (red) and $R=30$ (black). b)Plot of $\xi$ (solid) and $\dot{\eta}$ (dashed) vs time $t/\tau_{A}$ for $R=3$ (green), $R=10 \approx R^{\star}$ (blue), $R=16$ (red) and $R=30$ (black) to show that for $R\geq R^{\star}$, the outflows decelerate ($\dot{\eta}<\eta_0$) and for a high enough $R=30$, the outflows halt ($\dot{\eta}=0$).}
    \label{fig:crit_rc}
\end{figure}

The impact of these predicted values can clearly be seen in Figures \ref{fig:crit_rc}(a) and (b), which plot $\xi$ along with $\eta$ (a) and $\dot{\eta}$ (b) over time for $R=3$, (green), $R=10 \approx R^{\star}$ (red), $R= 16$ (blue) and $R=30$ (black) for bremsstrahlung cooling. In Figure \ref{fig:crit_rc}, it can be seen that the evolution of $\eta$ for $R=10 \approx R^{\star}$ (red) begins to deviate from the weak cooling case ($R=3$, green), whilst for $R=30$ (black), $\eta$ begins to collapse at $t<t_c$. In Figure \ref{fig:crit_rc}(b), the impact of cooling on the outflows can be seen more clearly. For weak cooling ($R=3)$, $\dot{\eta}$ always has a positive slope, however, in contrast, for $R\gtrsim 10 \approx R^{\star}$, $\dot{\eta}$ has some negative slope, indicating the deceleration of the outflows. Crucially, for $R=30$, which is the first value of $R$ that is distinctly greater than $R^{\star}$, the outflows are shown to not just decelerate, but to halt ($\dot{\eta}=0$) before $t_c$ and reverse in direction ($\dot{\eta}<0$).

\section{Modified Sweet-Parker Model}
\label{spmodel}
In reality, there will be mechanisms that stop the rapid collapse by heating the plasma and counteracting the strong inward thermal pressure gradient. Beyond compressional heating, a candidate for this is ohmic heating. The volumetric ohmic heating rate $\int \dot{Q}_{\Omega} dV$, where $V$ is taken to be the volume of the current sheet and $\dot{Q}_{\Omega}\equiv4\pi\chi_mj^2/c^2$, is deemed to be negligible as the magnetic diffusivity $\chi_m$ is small; however, close to $t_c$, the gradients in $B$ are strong enough to generate a large enough current $j$ such that $\dot{Q}_{\Omega}$ can no longer be neglected. This can be seen by evaluating the approximate time at which the resistive diffusion rate $\tau_{\chi_m}^{-1}\sim\chi_m/\xi^2$ and the X-point collapse rate along the $x$-axis $\tcol^{-1}\sim\Gamma_x=\dot{\xi}/\xi$ balance. Using the asymptotic expressions derived for $\xi$ in Equations \ref{eq:syro_scale} and \ref{eq:scale_largerc}, these terms balance at some time $t\sim t_c - (3\chi_m/2)^3\approx t_c$ for the case without cooling and $t\sim t_c -\chi_m \approx t_c$ with strong radiative cooling. Given this result, it is worth considering what a steady-state solution looks like, if one should exist. This can be achieved by expanding on the radiatively-cooled Sweet-Parker model developed by \citet{uzdensky2011magnetic} to include the length of the current sheet as a variable, as our current sheet model predicts that for strong enough cooling, the current sheet will have already started contracting before a steady-state is achieved.

For a current sheet of length $2L$ and width $2\delta$ such that $\delta\ll L$ (this remains true even if the current sheet contracts, see Figure \ref{fig:eta_xi_sol}), we define the upstream magnetic field, velocity, mass density and number density to be $\Bin$, $\vin$, $\rhoin$ and $\nin$, respectively. Similarly, we label the corresponding downstream magnetic field, velocity, mass density and number density to be $\Bout$, $\vout$, $\rhoout$ and $\nout$.  In contrast to the standard, incompressible Sweet-Parker model, there are two additional variables here: $\nout$ (or $\rhoout$) due to the inclusion of plasma compressibility, and $L$. The equations are closed by \citet{uzdensky2011magnetic} to find $\nout$ by assuming volumetric power balance between ohmic heating (which we approximate to be held locally) $\dot{Q}_{\Omega}=4\pi\chi_mj^2/c^2$ and radiative cooling $\dot{Q}_{\text{rad}}=\Ccool\rho^aT^b$. To close the equations with varying $L$, we further assume that this steady-state is achieved if the advection time $\tau_A$ is similar to the cooling time $\tau_{\text{cool}}$, allowing the reconnection process to occur before a given plasma parcel in the layer can radiatively collapse.

Starting with the MHD continuity equation, allowing for compressibility results in
\begin{equation}
    \rhoin \vin L \sim \rhoout \vout \delta \to \vin L \sim A \vout \delta,
\end{equation}
where we define the compression ratio $A\equiv \rho_X/\rhoin$. Here, for simplicity, we assume that the steady-state has no significant variation in density along the layer such that $\rho_X\approx \rhoout$ and $A\approx \rhoout/\rhoin$; this was also assumed in the calculation performed by \citet{uzdensky2011magnetic} and is justified by the presence of radiative cooling. Similar to the incompressible Sweet-Parker model, the steady-state assumption enforced via Ohm's law yields $\vin \sim \chi_m/\delta$. Additionally, the outflow component of the momentum equation can be used to find the outflow velocity: 
\begin{equation}
    \rhoout \vout^2/2 \sim \pin -P_{\rm out} + \frac{\Bin^2}{8\pi} + \frac{\Bin^2}{4\pi} A,
\end{equation}
where we have substituted $P_X$ using the pressure balance condition in the inflow direction $P_X=P_{\text{in}}+\Bin^2/8\pi$. 
We assume, as usual, that $\pin\approx P_{\rm out}$, implying $T_{\rm in}> T_{\rm out}$ for $A>1$, which is consistent with the radiative cooling of the current sheet. \footnote{It is worth noting that for this steady-state to exist, $P_X>\pin\approx P_{\rm out}$, implying an outward pointing thermal pressure gradient along the outflows. However, when the current sheet is collapsing, there exists a period where the pressure gradient along the outflows points inwards, resulting in the contraction of the length (as discussed in Section~\ref{sub:rcrit}). This reversal in the direction of the pressure gradient once the steady-state is established can be thought of as a consequence of the power balance imposed in the current sheet, resulting in ohmic heating counteracting radiative cooling losses.} 
This then yields
\begin{equation}
    \vout\approx V_A\left(\frac{1+2A}{A}\right)^{1/2},
\end{equation}
where the factor $(1+2A)^{1/2}/A^{1/2}$ can be shown to be of order unity (which we ignore for the purpose of this calculation) by considering the limits of $A$; for $A\sim\mathcal{O}(1)$, $\vout\approx \sqrt{3}V_A$ and for $A\gg1$, $\vout\approx \sqrt{2}V_A$.
\footnote{As also noted by \citet{uzdensky2011magnetic}, the outflow velocity here is Alfv\'enic rather than a hybrid velocity modified by the ratio $\rhoin/\rhoout\approx A^{-1}$ \citep{parker1963solar}, which is obtained by neglecting the MHD body force.}

Combining the expressions above, the following expression for the reconnection rate can be obtained:
\begin{equation}
    \frac{\vin}{V_A}\sim A^{1/2}S_L^{-1/2},
    \label{eq:recrate}
\end{equation}
where $S_L\equiv LV_A/\chi_m$ is the Lundquist number for a reconnection layer of length $L$. Thus, in order to make a definitive comparison between this expression and the expected incompressible Sweet-Parker reconnection rate in the absence of radiative cooling, explicit expressions for $A$ and $L$ need to be obtained by enforcing $\dot{Q}_{\Omega}=\dot{Q}_{\text{rad}}$ and $\tau_{\text{cool}}=\tau_A$. 

Starting with the power balance condition, which we approximate as $4\pi\chi_mj_X^2/c^2=\Ccool\rho_X^aT_X^b$, where the subscript `$X$' denotes quantities measured at the X-point, we have:
\begin{equation}
    \chi_m\frac{\Bin^2}{4\pi\delta^2}\sim \frac{\Ccoolbar}{\gamma-1}\rho_X^{a-b}P_X^b .
    \label{eq:pb}
\end{equation}
$P_X$ can be conveniently reexpressed in terms of $\Bin$ using the required pressure balance in the inflow direction $P_X=P_{\text{in}}+\Bin^2/8\pi=(1+\betain)\Bin^2/8\pi$, where $\betain\equiv P_{\text{in}}/(\Bin^2/8\pi)$ is the inflow plasma beta. It is also convenient to define $\mathcal{R}$ (Equation \ref{eq:fancyR}) for the inflows:
\begin{equation}
\mathcal{R}=\frac{\Ccoolbar\rhoin^{a-b}P_{\text{in}}^b}{\gamma P_{\text{in}} c_s/\Lsys}\approx\frac{\Ccoolbar\rhoin^{a-b}P_{\text{in}}^{b-1}\Lsys}{\gamma^{3/2}\betain^{1/2}V_A},
\end{equation}
where the sound speed is now defined in terms of inflow quantities $c_s=(\gamma P_{\text{in}}/\rhoin)^{1/2}$ and $\Lsys$ is the characteristic system size. 
Note that this is expressed entirely in terms of quantities assumed to be known.
We are specifically interested in cases where $\mathcal{R}\gg1$.

Plugging these expressions into Equation \ref{eq:pb} yields:
\begin{equation}
    \frac{L}{\Lsys}\sim \frac{(\gamma-1)}{\mathcal{R}\gamma^{3/2}}\frac{\betain^{b-3/2}}{(1+\betain)^b}A^{b-a+1}.
    \label{eq:pb_llsys}
\end{equation}

Next, enforcing $\tau_{\text{cool}}=\tau_A$ implies
\begin{equation}
    \frac{L}{V_A}=\frac{P_X}{\Ccool\rho_X^aT_X^b}=\frac{(\gamma-1)}{\Ccoolbar}P_X^{1-b}\rho_X^{b-a},
\end{equation}
which consequently yields the following expression for $L/\Lsys$ in terms of $A$, $\mathcal{R}$ and $\betain$:
\begin{equation}
    \frac{L}{\Lsys}\sim \frac{(\gamma-1)}{\mathcal{R}\gamma^{3/2}}\frac{\betain^{b-3/2}}{(1+\betain)^{b-1}}A^{b-a}
    \label{eq:tceqta}.
\end{equation}
Finally, equating the right-hand sides of Equation \ref{eq:pb_llsys} and Equation \ref{eq:tceqta} results in the following explicit expression for $A$:
\begin{equation}
    A\approx 1+\betain.
    \label{eq:a}
\end{equation}
This expression is notably independent of the cooling mechanism (at least as long as the radiative power loss due to the cooling mechanism can be parameterized as $\dot{Q}_{\text{rad}}=\Ccool\rho^aT^b$).
Unsurprisingly, we find that higher values of the upstream plasma beta result in larger compressibility factors.

The expression for $A$, in turn, can be used to obtain an expression for the length of the current sheet:
\begin{equation}
    L\sim \frac{(\gamma-1)}{\mathcal{R}\gamma^{3/2}} \frac{\betain^{b-3/2}}{(1+\betain)^{a-1}}\Lsys.
    \label{eq:l_lys}
\end{equation}
Consistent with our current sheet model, we find that $L<\Lsys$ when radiative cooling dominates compressional heating, $\mathcal{R}\gg 1$.

Finally, an explicit expression for the reconnection rate can be obtained by substituting Equations \ref{eq:a} and \ref{eq:l_lys} into Equation \ref{eq:recrate}:
\begin{equation}
       \frac{\vin}{V_A}\sim\frac{\mathcal{R}^{1/2}\gamma^{3/4}}{(\gamma-1)^{1/2}}\betain^{(3/2-b)/2}(1+\betain)^{a/2}S_{\Lsys}^{-1/2}.
       \label{eq:rec_rate_total}
\end{equation}
This expression is to be contrasted with the Sweet-Parker reconnection rate for an incompressible plasma with no radiative cooling, $\sim S_{\Lsys}^{-1/2}$. Thus, radiative cooling with $\mathcal{R}\geq1$ results in an increase in the reconnection rate --- although the dependence is relatively weak and not enough, \textit{per se}, to account for the values of the reconnection rates expected in the astrophysical environments where radiative cooling is a dominant effect. This higher reconnection rate is consistent with the statement that a larger $\betain$, and thus $A$, is expected in the presence of radiative cooling as the increased reconnection rate reduces the pile-up of flux at the inflows, resulting in a lower inflow magnetic pressure $\Bin^2/8\pi$ \citep{Priest_1986,Datta_jpp_2024}.

For completeness, the expression obtained from Ohm's law $\vin \sim \chi_m/\delta$ and Equation \ref{eq:rec_rate_total} can now be used to yield an explicit expression for the current sheet thickness $\delta$:
\begin{equation}
\delta \sim \frac{(\gamma-1)^{1/2}}{\mathcal{R}^{1/2}\gamma^{3/4}}\betain^{(b-3/2)/2}(1+\betain)^{-a/2} \delta_{\text{SP}}
\end{equation}
where $\delta_{\text{SP}}\sim \Lsys S_{\Lsys}^{-1/2}$ is the expected current sheet width for incompressible, non-radiative Sweet-Parker reconnection.

Lastly, specific expressions for bremsstrahlung cooling can be obtained by setting $a=2, b=1/2$:
\begin{equation}
L\sim\frac{(\gamma-1)}{\mathcal{R}\gamma^{3/2}}\frac{1}{\betain(1+
    \betain)}\Lsys,
    \label{eq:llsys}
\end{equation}
\begin{equation}
\frac{\vin}{V_A}\sim\frac{\mathcal{R}^{1/2}\gamma^{3/4}}{(\gamma-1)^{1/2}}\betain^{1/2}(1+\betain)S_{\Lsys}^{-1/2},
    \label{eq:reconrate}
\end{equation}
\begin{equation}
    \delta\sim\frac{(\gamma-1)^{1/2}}{\mathcal{R}^{1/2}\gamma^{3/4}}\frac{\delta_{\text{SP}}}{\betain^{1/2}(1+\betain)},
\end{equation}
implying that a steady-state exists for a reconnection layer cooled by bremsstrahlung radiation. This is in contrast to the prediction made by \citet{uzdensky2011magnetic}, who predict that bremsstrahlung cooling is not thermally stable, and will lead to a thermal cooling catastrophe in the layer. The stability condition that yields this prediction is derived for $A\gg1$, under the assumption that the initial state is a standard Sweet-Parker current sheet that evolves under radiative cooling. Here, the current sheet forms via the X-point collapse model we derive, which requires us to assume $L$ to be a variable, allowing us to find a steady-state solution.
\section{Conclusions}
\label{discussion}
In this paper, we develop a simple, but analytically solvable, model for current sheet formation through X-point collapse in the presence of radiative cooling. This is achieved by extending the calculation performed by \citet{Syrovatskii_1971} to include an optically thin radiative cooling term $\dot{Q}_{\text{rad}}\sim \Ccool\rho^aT^b$ in the MHD equation of state. The resultant equations (\ref{final_rcool_eq}) for $\xi(t)=x(t)/x_0$ and $\eta(t)=y(t)/y_0$ for the location of a plasma parcel originally at position ($x_0$,$y_0$) are derived using Lagrangian MHD (see Sections \ref{sub:lmhd}-\ref{sub:leom_rc}) and account for the strength of radiative cooling using the cooling parameter $R=\tout/\tau_{\text{cool}}$. 

The solutions to these equations are obtained numerically for bremsstrahlung cooling with values of $R$ ranging from $R=3$ (weak cooling) to $R=3000$ (strong cooling), and are compared to the solution with uniform pressure and no radiative cooling, as  obtained by \citet{Syrovatskii_1971}, and the solution with a non-uniform pressure and no radiative cooling (see Section \ref{sub:numerical}). The results, which can be seen by plotting the evolution of $\xi$ and $\eta$ (Figure \ref{fig:eta_xi_sol}) and the contours of the magnetic flux function $\psi (x,y)$ (Figure \ref{fig:xpointevolution}), show that stronger radiative cooling causes the X-point to collapse more rapidly in the inflow direction, and also results in a reversal of the direction of the outflows for sufficiently large values of $R$.

The derived asymptotic behavior of $\xi$ (see Section \ref{subsec:asymptotics}) can be used to show that, for weak cooling, the radiative cooling term adds a corrective term that scales as $\sim (t_c-t)^{4/3}$ to the non-radiative asymptotic behavior $\xi \sim (t_c-t)^{2/3}$. For strong cooling, the cooling term dominates the asymptotic behavior of $\xi$ (Figure \ref{fig:individual_terms}(b)) resulting in a modified scaling of $\xi \sim (t_c-t)$, which has a sharper scaling than that for weak cooling.

The reversal of flows along the $y$-axis can be attributed to the non-uniform compression of the layer by the magnetic pressure, resulting in an inward thermal pressure gradient that opposes the $\mathbf{j} \times \mathbf{B}$ force that drives the kinetic pressure gradient formed by the outflows. Naturally, a stronger $\mathbf{j} \times \mathbf{B}$ (stronger upstream magnetic field) yielding faster outflows reduces the impact of the inward thermal pressure gradient, and thus radiative cooling. The threshold for when this inward thermal pressure gradient is strong enough to decelerate the outflows is found to occur above a critical value of $R\equiv R^{\star} \approx 10$. At $R^{\star}$, the cooling term in the equation for $\ddot{\xi}$ balances the $1/\xi^2$ term near the finite-time singularity (see Section \ref{sub:rcrit}). 

Given the sharp magnetic field gradient across the width of the current sheet for $t\approx t_c$, ohmic heating is expected to no longer be negligible around this time. Thus, we derived a steady-state solution for the radiatively-cooled current sheet by modifying the compressible Sweet-Parker model derived by \citet{uzdensky2011magnetic}.
Importantly, our modification includes the current sheet length as a variable, as our current sheet formation model predicts that, for strong cooling, the expected current sheet length will be shorter than the characteristic system size. This additional variable is solved for by imposing the condition $\tau_A=\tau_{\text{cool}}$, allowing for a given plasma parcel to travel the length of the layer before undergoing radiatively driven collapse (see Section \ref{spmodel}); this condition is required for the existence of a steady-state. It is found that the radiative cooling results in an increased reconnection rate as compared to the Sweet-Parker reconnection rate for an incompressible plasma (Equation \ref{eq:reconrate}). Effectively, radiative cooling yields a contracted steady-state current sheet length compared to the system size (Equation \ref{eq:llsys}), which in turn decreases the Lundquist number and thus results in a higher reconnection rate. 

It is worth noting that whilst this paper focuses on radiative cooling, radiation can have a myriad of important effects on magnetic reconnection in extreme astrophysical environments such as radiation pressure, Compton drag and, for extremely high energy densities, pair creation \citep{uzdensky2011magneticb}.
In addition to this, reconnection layers in higher energy density environments can be optically thick to their dominant cooling mechanism, requiring the inclusion of the radiative transfer equations in the reconnection problem. Given the vast variety of processes that can be accounted for when modeling the interaction between radiation and reconnection, we have chosen to start by considering the simplest set up, involving the dominant effect of radiation, radiative cooling, that is optically thin in the relatively low energy density plasmas for which the MHD framework is valid. Thus, influenced by the early progress made in modeling reconnection in the classical setting, the work in this paper provides a starting point for the development of a theoretical framework for reconnection processes in radiatively-cooled plasmas. The findings in this paper lay the groundwork for future study on the impact of radiative cooling on reconnection beyond the Sweet-Parker model, including its interaction with the plasmoid instability and its impact on the global reconnection rate. Interesting directions for future research include numerical verification of the results in this paper and a study on the inclusion of other important thermodynamic effects such as thermal conduction.

\section{Acknowledgments}
The authors would like to thank Jack Hare and Dmitri Uzdensky for valuable discussions on radiative cooling in reconnection layers. SC acknowledges support from the MIT MathWorks and the NSE Manson Benedict fellowship awards.

\section{Declaration of Interests}
The authors have no conflicts of interest to disclose.

\appendix

\section{}\label{appA}
Here, we detail the method used to obtain the final form of the Lagrangian pressure $P_L$ in Equation \ref{eq:lagrangian_pressure}.

Starting from the Lagrangian equation of state described by Equation \ref{eq:lagrangian_eos}:
\begin{equation}
\frac{\partial}{\partial t} \ln \left[\frac{P_L}{\rho_L^\gamma}\right]=-\Ccoolbar \rho_L^{a-b} P_L^{b-1},
\end{equation}
we define an ansatz $P_L\rho_L^{-\gamma}=F(t,x,y)P_0\rho_0^{-\gamma}$ with $F(t,x,y=0)=1$. To substitute this into the equation of state, we first rearrange the right hand side term of the equation of state as follows:
\begin{equation}
    \frac{P_L^{b-1}}{\rho_L^{b-a}}=\frac{P_L^{b-1}}{\rho_L^{\gamma(b-1)} \rho_L^{b-a} \rho_L^{-\gamma(b-1)}}.
\end{equation}
Using the expression for Lagrangian density $\rho_L=\rho_0 / J$, the following expression for $\rho_L^{a-b} P_L^{b-1}$ in terms of $F(x_0,y_0)$, $P_0$, $\rho_0$ and $J(t,\boldsymbol{r}_0)$ is obtained:
\begin{equation}
 \frac{P_L^{b-1}}{\rho_L^{b-a}}=F(x_0,y_0)^{b-1} \frac{P_0^{b-1}}{\rho_0^{b-a}} J(t,\boldsymbol{r}_0)^{b(1-\gamma)+(\gamma-a)}.
 \label{apdx:eq:pl_rho}
\end{equation}
This expression thus encodes the time dependence of the left hand side of the equation into $F(x_0,y_0)$, which we have to solve for, and $J(t,\boldsymbol{r}_0)$, which is evaluated to be $J(t,\boldsymbol{r}_0)=\xi\eta$ in Equations \ref{final_rcool_eq}. By simply substituting the expression in Equation \ref{apdx:eq:pl_rho} in the equation of state, the following partial differential equation in $F$ can be found:
\begin{equation}
\frac{\partial}{\partial t} \ln \left[F \frac{P_0}{\rho_0^\gamma}\right]=\frac{1}{F^{b-1}} \frac{\partial}{\partial t} \ln F=\frac{1}{F^b}\frac{\partial F}{\partial t}=-C_{t}\times F^{b-1} P_0^{b-1} \rho_0^{a-b} J^{b(1-\gamma)+(\gamma-a)}.
\end{equation}
Combining the homogeneous solution $C_1$, which is a constant to be determined, with the particular solution that integrates over $J(t,\boldsymbol{r}_0)$ results in the following general solution for $F$:
\begin{equation}
     F=C_1-\frac{C_2(x_0, y_0)}{P_0}[(\gamma-1)(1-b)\Ccoolbar\int_0^t J(t',\boldsymbol{r}_0)^{b(1-\gamma)+(\gamma-a)}dt^{\prime}]^{1/(1-b)}.
\end{equation}
To completely specify $F$, the constants $C_1$ and $C_2$ need to be determined. $C_1$ can be evaluated using the initial condition $F(t=0,x_0,y_0)=1$ to simply be $C_1=1$. On the other hand, $C_2(x_0,y_0)$ needs to be selected;  we define it to be $P_{0}=P_{0,X}\left(1-(x_0^2+y_0^2)/2\right)$. This expression is just a parabolic, non-uniform pressure profile used to model the plasma in this paper. It is noted that $C_2(x_0,y_0)$ can take any spatial dependence given that the initial condition for $F$ is satisfied regardless of the form of $C_2(x_0,y_0)$.

Finally, substituting this expression for $F$ into the ansatz $P_L\rho_L^{-\gamma}=F(t,x_0,y_0)P_0\rho_0^{-\gamma}$, with the expression for $\Ccooltilde=\left[(\gamma-1)(b-1) R\right]^{1/(1-b)}$ yields the expression required for $P_L$:
\begin{equation}
P_L=\frac{P_0\left(x_0, y_0\right)}{J^\gamma} \times\left(1-\left[(\gamma-1)(1-b) R\int_0^t J^{b(1-\gamma)+(\gamma-a)}\left(t',\boldsymbol{r}_0\right) d t^{\prime}\right]^{1 /(1-b)}\right),
\end{equation}
where $R\equiv\tout/\tau_{\text{cool}}$ is the cooling parameter.

\section{}\label{appB}
Here, we justify and detail the method used to obtain the correction $\sim(t_c-t)^{4/3}$ to scaling for $\xi$ for small values of $R$.
First, we justify the use of our method by considering the alternative method employed in S71, starting with the equation obtained for $\ddot{\xi}$ (Equation \ref{eq:diffeq_xi_smallrc}), which is expressed in the limit of small $R$, near $t\approx t_c$:
\begin{equation}
    \ddot{\xi} \approx-\eta\left(t_c\right)\left[\frac{1}{\xi^2}+C_{\text{int}}(t_c-t)^{\frac{2}{3} \alpha_p+\frac{1}{(1-b)}}\right]; \quad \text{with } \alpha_p=\frac{b-a}{(1-b)},
\end{equation}
where $C_{\text{int}}$ is:
\begin{equation}
    C_{\text{int}}=\Ccooltilde \left[\frac{9}{2} \eta(t_c)^4\right]^{(b-a)/ 3(1-b)}\left[\frac{3}{2[b(1-\gamma)+(\gamma-a)]+3}\right]^{1/(1-b)}.
\end{equation}
Here, we are considering weak enough cooling such that there is no turning point in $\eta(t)$ and that $1/\xi^2\gg1/\eta^2$. As such, we take $\eta\approx \eta(t_c)$ and neglect the $1/\eta^2$ and $1/(\xi\eta)^{\gamma}$ terms ($1/\xi^2\gg1/(\xi\eta)^{\gamma}$ as $\gamma=5/3 <2$).

For bremsstrahlung radiation ($a=2,b=1/2)$, this yields the following second order ordinary differential equation for $\xi$:
\begin{equation}
    \ddot{\xi} \approx -\eta(t_c)\left(\frac{1}{\xi^2}+C_{\text{int}}\right)
    \label{appdx:eq:ddotxi_brems}
\end{equation}
In S71, this second order ODE is reduced to a first order ODE by substituting the expression for $\ddot{\xi}$ in the following taylor expansion, taken in the limit $t\approx t_c$:
\begin{equation}
    \xi(t_c)=0 \approx \xi(t)+(t_c-t)\dot{\xi}+\frac{(t_c-t)^2}{2}\ddot{\xi}+\mathcal{O}\left((t_c-t)^3\right).
    \label{appdx:eq:taylor_expansion}
\end{equation}
Defining $(t_c-t)\equiv s$, this yields the following first order ODE for $\xi$ in $s$:
\begin{equation}
    \xi^3-\xi^2s\frac{d\xi}{ds}-\eta(t_c)C_{\text{int}}\xi^2s=\eta(t_c)s.
    \label{appdx:eq:xi_ode}
\end{equation}

This equation can be identified as a form Chini's equation; properties of this can be used to show that an analytic solution to this equation does not exist. Analytic solutions are possible for such equations if their corresponding Chini invariant ($I_{ch}$, defined below) is independent of the variable (\cite{Kamke_1944b}). For the general form of Chini's equation for some function $y(x)$:
\begin{equation}
    \frac{dy(x)}{dx}=f(x)y(x)^n-g(x)y(x)+h(x),
\end{equation}
the expression for the Chini invariant $I_{ch}$ is defined as follows:
\begin{equation}
    I_{ch}=f(x)^{-(n+1)}h(x)^{-2n+1}\left(f(x)\frac{dh(x)}{dx}-h(x)\frac{df(x)}{dx}-n\times g(x)f(x)h(x)\right)^n \times n^{-n}.
\end{equation}
Our differential equation can be reexpressed to match the general form of Chini's equation:
\begin{equation}
    \frac{d\xi}{ds}=\frac{\xi}{s}-C_{\text{int}}\eta(t_c)-\frac{\eta(t_c)}{\xi^{2}},
\end{equation}
yielding a corresponding Chini invariant:
\begin{equation}
    I_{ch}=[C_{\text{int}}\eta(t_c)^2]^{-3}\times\left(-2\times C_{\text{int}}\eta(t_c)^2\times\frac{1}{s}\right)^{-2}\times(-2)^{2}=[C_{\text{int}}\eta(t_c)^2]^{-5}s^2.
\end{equation}
Here, $I_{ch}$ is clearly a function of the variable $s$ and thus there is no analytic solution for Equation \ref{appdx:eq:xi_ode}. Thus, we choose to proceed using a different method, that employs the use of the Taylor expansion in Equation \ref{appdx:eq:taylor_expansion} at a later step. 

Returning to Equation \ref{appdx:eq:ddotxi_brems} for bremsstrahlung radiation, we can use the substitution $\xi\equiv c_1^{1 / 3} z $ to yield the simpler equation:
\begin{equation}
   \ddot{z}=\frac{1}{z^2}+\frac{c_2}{c_1^{1 / 3}},
\end{equation}
where $c_1=-\eta(t_c)$ and $c_2=-C_{\text{int}} \eta(t_c)$. Then, by simply integrating by parts, we can reduce this equation to the following first order ODE:
\begin{equation}
    \frac{1}{2}\left(\dot{z}\right)^2=c_3+\frac{c_2}{c_1^{1/3}} z-\frac{1}{z},
\end{equation}
where $c_3=c_1^{1 / 3}(c_1^{-1 / 3}/2-c_2c_1^{2 / 3})$. 
By rearranging this ODE and integrating over time, a solution in the functional form $\Delta t(z)$, where $\Delta t=(t_c-t)$, can be obtained:
\begin{equation}
    \left(t_c-t\right)+c_4=\int^{\xi(t_c)}_{\xi_0} \sqrt{\frac{\xi}{c_3 c_1^{2 / 3} \xi+c_2 \xi^2-c_1} }d \xi, 
\end{equation}
where $c_4=0$ as $\xi(t=t_c)=0$.
Treating $\xi$ as the relevant small parameter, the integrand can be expanded to yield:
\begin{equation}
 \begin{aligned}
    \left(t_c-t\right) &\approx \int \sqrt{-\frac{1}{c_1} \xi}-\frac{1}{2} \xi^{3/2}\left(c_3 c_1^{2/3} \sqrt{-\frac{1}{c_1}}\right)+\mathcal{O}\left(\xi^{5/2}\right) d \xi \\
    & \approx \frac{2}{3} \sqrt{-\frac{1}{c_1}} \xi^{3/2}-\frac{1\left(c_3 c_1^{2/3} \sqrt{-\frac{1}{c_1}}\right)}{5} \xi^{5/2}+\mathcal{O}\left(\xi^{7/2}\right). \\
 \end{aligned}
\end{equation}
Lastly, substituting the ansatz $\xi \sim (9\eta(t_c)/4)^{1/3}(t_c-t)^{2/3}+c_\alpha(t_c-t)^{\alpha} $, and keeping terms up to $\mathcal{O}(\xi^{5/2})$ gives $\alpha=4/3$, yielding the required scaling for $\xi$:
\begin{equation}
    \xi \sim \left(\frac{9}{4}\eta(t_c)\right)^{1/3}(t_c-t)^{2/3}+\frac{2}{15}\left(\frac{9}{4}\right)^{13/6}c_3(-c_1)^{7/3}(t_c-t)^{4/3}.
    \label{eq:weakcool_scale}
\end{equation}

\bibliographystyle{jpp}

\bibliography{paper.bib}

\end{document}